\documentclass[preprint]{emulateapj}

\shorttitle{True Masses of Exoplanets} 
\shortauthors{Batygin \& Laughlin} 

\begin{document}
 
\title{Resolving the Sin(I) degeneracy in Low-Mass Multi-Planet Systems}  
\author{Konstantin Batygin$^1$ \& Gregory Laughlin$^2$} 

\affil{$^1$Division of Geological and Planetary Sciences, California Institute of Technology, Pasadena, CA 91125} 
\affil{$^2$UCO/Lick Observatory, Department of Astronomy \& Astrophysics, University of California at Santa Cruz, Santa Cruz, CA 95064} 

\email{kbatygin@gps.caltech.edu}
 
 \begin{abstract}
Long-term orbital evolution of multi-planet systems under tidal dissipation often converges to a stationary state, known as the tidal fixed point. The fixed point is characterized by a lack of oscillations in the eccentricities and apsidal alignment among the orbits. Quantitatively, the nature of the fixed point is dictated by mutual interactions among the planets as well as non-Keplerian effects. We show that if a roughly coplanar system hosts a hot, sub-Saturn mass planet, and is tidally relaxed, separation of planet-planet interactions and non-Keplerian effects in the equations of motion leads to a direct determination of the true masses of the planets. Consequently, a ``snap-shot" observational determination of the orbital state resolves the $\sin(I)$ degeneracy, and opens up a direct avenue towards identification of the true lowest-mass exo-planets detected. We present an approximate, as well as a general, mathematical framework for computation of the line of sight inclination of secular systems, and apply our models illustratively to the 61 Vir system. We conclude by discussing the observability of planetary systems to which our method is applicable and we set our analysis into a broader context by presenting a current summary of the various possibilities for determining the physical properties of planets from observations of their orbital states.

 \end{abstract}
 
 
 \section{Introduction}

Since the seminal discovery of the first giant planet orbiting a main sequence star (Mayor \& Queloz 1995), using the radial velocity (RV) method, over 400 additional extra-solar planets have been confirmed. The greatest disadvantage of the RV method lies in the uncertainty of the true masses of the discovered planets, as the inclination of the orbits to the line of sight, $I$, are unknown. In resonant systems, such as GL876, monitoring of the resonant argument and the precession rates may lead to determination of the true masses (e.g. Rivera et al 2005). In the vast majority of cases, however, the $\sin(I)$ degeneracy, remains a continued source of frustration.

Still, RV surveys persist in yielding fruitful results, and the continued detection of exo-planets has brought forth many surprises. Perhaps one of the biggest surprises has been the discovery of extremely close-in bodies whose mass-range spans the entire planetary spectrum. These objects have since become a subject of fascination in the community and more importantly, have provided a new test-bed for various theoretical efforts. 

Extra-solar multi-planet systems that host ``hot" planets differ drastically from our own solar system in many ways, including orbital dynamics. In our solar system, gravitational interactions among the planets are sufficient to, at least approximately, explain orbital evolution. In many extra-solar planetary systems however, similarly to the case of the Galilean satellites, dissipation of orbital energy due to tides plays an unavoidably important role. The long-term effect of this additional interaction provides an opportunity to infer important additional properties of the system that cannot be observed directly.

Qualitatively speaking, in a system of two or more planets  that are not in a mean-motion resonance and are roughly coplanar, tides drive the orbits towards a stationary state i.e. a ``fixed point".  A fixed point is characterized by continued apsidal alignment and a well-determined eccentricity ratio that is nearly constant in time (Wu \& Goldreich 2002, Mardling 2007). The factors that determine the actual quantitative nature of the state are not limited to gravitational planet-planet interactions. Indeed, general relativistic and tidal corrections, among other things, play a crucial role. It is through these ``non-Keplerian" interactions that additional information can be learned, as they are governed by parameters other than just planetary masses,.

Upon discovery of the first multiple planetary system with a transiting ``hot Jupiter", Hat-P-13 (Bakos et al 2009), it was pointed out that the system likely resides at a fixed point (Batygin, Bodenheimer \& Laughlin 2009). Furthermore, it was shown that as the mass and radius of the inner planet are known, consideration of the planetary quadru-pole gravitational field, and its contribution in determination of the fixed point leads to a direct measurement of the planetary interior structure. In other words, a precise ``snap-shot" of the orbits gives the planetary Love number, $k_2$, which is a measure of the interior density distribution, with high accuracy.  

The last decade of observations has revealed that generally, hot Jupiters tend not to be accompanied by readily detectable companion planets (Ragozzine \& Holman 2010). Smaller planets, such as hot Neptunes and hot Super-Earths, however, tend to occur in multiple-planet systems (Lo Curto et al 2010), hinting at different migration histories (Terquem \& Papaloizou 2007). 

Here, we consider the latter class of systems, with an eye toward inferring conventionally unobservable planetary properties that influence the details of the fixed-point configuration. In particular, we show that if a non-transitng (RV) system hosts a small ($R \lesssim R_{\textrm{Nep}}$) hot planet, it is possible to derive the true masses of the planets i.e. resolve the $\sin(I)$ degeneracy from a detailed determination of the system's orbital state. The plan of the paper is as follows: in section 2, we outline our mathematical model. In section 3, we apply the theory to the 61 Vir (Vogt et al 2010) system. In section 4, we discuss the possibility of determination of the radius and interior structure of massive RV planets. We conclude and discuss our results in section 5.

\section{Dynamical Evolution of a Planetary System With a Close-in Planet}

As already mentioned above, there are important differences between the dynamics of systems with and without close-in planets. In conservative (Hamiltonian) systems, of which our solar system provides an excellent approximation (Laskar 2008), the motion of the planets is subject to Liouville's theorem. Accordingly, strictly Hamiltonian flow can have no attractors in phase-space (Morbidelli 2002). Conversely, in dissipative systems, the phase space volume explored by the system continuously contracts, and truly steady-state solutions are possible. In other words, tides are needed for the system to arrive to a stationary state.

The path that the system will take to the fixed point is non-unique and depends on the initial conditions. Consequently, the initial transient period will also depend on the initial state. However, the fixed-point itself is unique for a chosen set of system parameters, and the system has no memory of its own evolution once it arrives to the fixed point. Thus, any quantity that is inferred from the fixed point is independent of the system's formation history. We now describe a mathematical model for the system's evolution to a stationary state and its orbital characteristics.

\subsection{Secular Interactions WIth non-Keplerian Effects}

Whenever planets are far away from low-order mean motion commensurabilities and the orbits are not changing significantly on the orbital time-scale (i.e. planets are not scattering), a secular approximation to the dynamics can be made. The secular approximation refers to an averaging procedure, where the gravitational potential between planets is averaged over the mean longitudes, thereby reducing the degrees of freedom inherent to the problem.  

Since the pioneering work of Laplace (1772) and Lagrange (1776), a number of perturbation theories based on various approximation of the disturbing potentials have been developed and applied in both solar system and exoplanetary contexts (Le Verrier 1856, Brouwer \& van Woerkom 1950, Laskar 1986, Laskar 2008, Eggleton \& Kiseleva-Eggleton 2001, Mardling \& Lin 2002, Lee \& Peale 2003, Michtchenko \& Malhotra 2004, Migaszewski \& Go{\'z}dziewski 2009, Lovis et al 2010, etc). Still, it is perhaps easiest to illustrate the ideas presented here in the context of a modified Laplace-Lagrange (LL) secular theory. 

The classical secular disturbing function (planet-planet potential), of N secondaris that interact solely by Newtonian gravity, expanded to first order in masses and second order in eccentricities reads (Murray \& Dermott 1999)
\begin{equation}
\mathcal{R}_{j}^{(sec)} =n_{j}a_{j}^{2} [\frac{1}{2} A_{jj} e_{j}^{2} + \sum_{k=1,k\neq{j}}^{N} A_{jk}e_{j}e_{k}\cos(\varpi_{j}-\varpi_{k}) ]
\end{equation}
where $e$ is eccentricty, $\varpi$ is the longitude of perihelion, $a$ is semi-major axes and $n$ is mean motion. The constant coefficients $A$ take the form
\begin{equation}
A_{jj}=\frac{n_{j}}{4}\sum_{k=1,k\neq{j}}^{N}\frac{ {m_{k}} } {M_{\star}+m_{j}}\alpha_{jk} \bar{\alpha}_{jk} b_{3/2}^{(1)}(\alpha_{jk})
\end{equation}
\begin{equation}
A_{jk}=-\frac{n_{j}}{4}\frac{ {m_{k}} } {M_{\star}+m_{j}}\alpha_{jk} \bar{\alpha}_{jk} b_{3/2}^{(2)}(\alpha_{jk})
\end{equation}
where $\alpha_{jk} = a_{j}/a_{k}$ if $( a_{j} < a_{k})$; $ a_{k}/a_{j}$ if $(a_{k} < a_{j})$, $\bar{\alpha}_{jk} = \alpha_{jk}$ if $( a_{j} < a_{k})$; $ 1$ if $(a_{k} < a_{j})$, $b_{3/2}^{(1)}(\alpha_{jk})$ \& $b_{3/2}^{(2)}(\alpha_{jk})$ are Laplace coefficients of first and second kind respectively, and $m = \tilde{m}/\sin(I)$ are the true masses of the planets i.e. $ \tilde{m}$ are the measured minimum masses and $I$ is the inclination of the system from line of sight.

Upon application of the linear form of Lagrange's planetary equations in terms of polar coordinates ($h= e \cos(\varpi),k= e \sin(\varpi)$), a linear system of ODE's emerges, where the $\bf{A}$ matrix encapsulates the dynamics of the system:
\begin{eqnarray}
\frac{dh_j}{dt}=\sum_{k=1}^{N} A_{jk} k_{k} \ \ \ \ \ \frac{dk_j}{dt}=-\sum_{k=1}^{N} A_{jk} h_{k}
\end{eqnarray}
We can express the system of equations more compactly by switching to complex Poincar\'e variables $z \equiv e \ \bold{e}^{i \varpi} = h + i k$. Simple chain rule yields
\begin{equation}
\frac{dz_j}{dt}=\sum_{k=1}^{N} i A_{jk} z_{jk} 
\end{equation}
This eigensystem can be solved in the standard way, similar to the problem of N coupled pendulums, and the solution reads:
\begin{equation}
z_{j}(t) = \sum_{k}^{N}  \beta_{jk} \ e^{i(g_{k}t + \delta_{k})} 
\end{equation}
where $g$'s are the eigenfrequencies and $\beta$'s are the eigenvectors of the $\bf{A}$ matrix. The relative amplitudes of the eigenvectors and the corresponding phases, $\delta$,  are determined by initial conditions. 

The above formulation does not take into account the additional orbital precession induced by general relativity (GR), stellar and planetary spin, and the tidal bulges of the star and the planet. The classical LL solution often gives poor quantitative approximations to the orbital evolution of extra-solar planets (Veras \& Armitage 2007), where the additional precession can dominate (Ragozzine \& Wolf 2009). 

The contributions to apsidal precession from the above-mentioned effects can be written as follows (Sterne 1939):
\begin{equation}
\left( \frac{d \varpi}{dt} \right)_{GR} = \frac{3G M_{\star} n}{a c^2 (1-e^2)}
\end{equation}
\begin{eqnarray}
\left( \frac{d \varpi}{dt} \right)_{spin} = \frac{n \ k_{2 p}}{2 (1-e^2)^{2}} \left( \frac{R_{p}}{a} \right)^5 \left( \frac{\Omega_{p}^2 a^3}{G m_{p}} \right) \nonumber \\ 
 + \frac{n \ k_{2 \star}}{2 (1-e^2)^{2}} \left( \frac{R_{\star}}{a} \right)^5 \left( \frac{\Omega_{\star}^2 a^3}{G M_{\star}} \right)
\end{eqnarray}
\begin{eqnarray}
\left( \frac{d \varpi}{dt} \right)_{tidal} &=& \frac{15 n}{2} k_{2p} \left( \frac{R_p}{a} \right)^5 \frac{M_{\star}}{m_p} \left( \frac{1+ \frac{3}{2}e^2+ \frac{1}{8}e^4}{(1-e^2)^5} \right) \nonumber \\ 
&+& \frac{15 n}{2} k_{\star} \left( \frac{R_{\star}}{a} \right)^5 \frac{m_{p}}{M_{\star}} \left( \frac{1+ \frac{3}{2}e^2+ \frac{1}{8}e^4}{(1-e^2)^5} \right)
\end{eqnarray}
where $c$ is the speed of light, $k_2$ is the Love number (twice the apsidal motion constant), $R$ is physical radius and $\Omega$ is the spin frequency. In equations (8) and (9) the first terms correspond to the planet and the latter terms correspond to the star. Neglecting higher-order effects, the total additional apsidal precession, evaluated for each planet can be organized into a square diagonal matrix
\begin{equation}
B_{jj}= \left( \frac{d \varpi}{dt} \right)_{GR} + \left( \frac{d \varpi}{dt} \right)_{spin} + \left( \frac{d \varpi}{dt} \right)_{tidal},
\end{equation}
and added to the $\bf{A}$ matrix in equation (5). This matrix is not to be confused with the mutual inclination interaction matrix (see Murray \& Dermott 1999), for which the standard notation is the same. When evaluating the additional precessions (the $\bf{B}$ matrix) in the context of LL theory, it is customary to expand equations (7) - (9) to first order in $e$, such that the dependence on $e$ disappears, and equation (5) remains linear in eccentricity, thus retaining its analytical solution. The augmentation of the diagonal matrix coefficients will modify the eigensystem quantitatively. However, the qualitative essence of the solution remains unchanged: the solution (equation 6) is still a sum of sinusoids with constant amplitudes.

So far, we have retained all additional precession terms for the sake of completeness. Before proceeding further, let us examine the relative importance of the terms that depend on the physical properties of the bodies with respect to GR, which is a purely geometrical effect and only depends on stellar mass and the orbital parameters. Consider the following dimension-less numbers
\begin{equation}
\Lambda_{spin}^{p} = \frac{c^2k_{2p} R_{p}^5 \Omega_p^2}{6 a G^2 M m_p } \ \ \ \ \ \Lambda_{spin}^{\star} = \frac{c^2 k_{2\star} R_{\star}^5 \Omega_{\star}^2}{6 a G^2 M_{\star}^2 } 
\end{equation}
\begin{equation}
\Lambda_{tidal}^{p} = \frac{5 c^2 k_{2p} R_p^5}{2 a^4 G m_{p}} \ \ \ \ \ \Lambda_{tidal}^{\star} =  \frac{5 c^2 k_{2\star} R_{\star}^5 m_{p}}{2 a^4 G M_{\star}^2}.
\end{equation}
A Jupiter-like planet at a characteristic close-in orbit ($P \sim 3$ days) has $\Lambda_{spin}^p \sim 0.05$ $\Lambda_{spin}^{\star} \sim 3 \times 10^{-5}$, $\Lambda_{tidal}^p \sim 1$ and $\Lambda_{tidal}^{\star} \sim 5 \times 10^{-4}$. Inflated hot Jupiters will often have $\Lambda_{tidal}^p \gg 1$, due to the $R_p^5$ dependence of the tidal term. Thus, precession rates of many hot Jupiters are completely dominated by the planetary tidal term, distantly followed by GR. As mentioned above already, this effect has has been used to infer the interior structures of transiting hot Jupiters, both in isolation (Ragozzine \& Wolf 2009) and in the presence of a perturbing companion (Batygin, Bodenheimer \& Laughlin 2009). Conversely, for a Neptune-like planet on a 3-day orbit, $\Lambda_{spin}^p \sim 0.005$ $\Lambda_{spin}^{\star} \sim 3 \times 10^{-5}$, $\Lambda_{tidal}^p \sim 0.1$ and $\Lambda_{tidal}^{\star} \sim 3 \times 10^{-6}$. The numbers continue to decline for super-Earths and terrestrial planets. This implies that in practice, the apsidal advance, resulting from rotation of both the planet and the star, as well as that resulting from the stellar tidal bulge, can often be neglected. Indeed, the situation is bimodal: for large planets, tidal precession dominates, where as for small planets, GR dominates the extra apsidal advance.

Let us now add dissipative tides to the system. Generally, tidal heating conserves the total angular momentum, but not energy. This leads to decay of the planetary eccentricity, as well as decay (or growth, depending on stellar spin) of the planet's semi-major axis (Goldreich 1963). The evolution of the semi-major axis happens over a much longer time-scale than that of the eccentricity, so in our simplified model, we adopt the standard practice of parameterizing tides with a constant decay of the eccentricity, $dz/dt=z/\tau_{c}$, where $\tau_{c}$ is the circularization timescale (Goldreich \& Soter 1966):
\begin{equation}
\tau_{c}=\frac{P_{p}}{21 \pi} \frac{Q_{p}}{k_{p}} \frac{m_{p}}{M_{\star}} \left(\frac{a}{R_{p}}\right)^{5}.
\end{equation}
Here, $P$ is the orbital period and $Q$ is a tidal quality factor. In a similar fashion as above, each planet can be subjected to tidal damping of eccentricity by constructing a square diagonal matrix with the elements $C_{jj} = 1/{\tau_{c}^{(j)}}$. Note that because tidal dissipation only affects semi-major axes, eccentricities and rotation rates directly, an identical procedure cannot be carried out for the mutual inclination eigenmode solution (see Mardling 2010 for an in-depth discussion). The equation of motion that accounts for the additional precession and tidal damping of the eccentricity takes the form:
\begin{equation}
\frac{dz_j}{dt}=\sum_{k=1}^{N} \left[ i (A_{jk}+B_{jk}) z_{jk} + C_{jk}  z_{jk} \right]
\end{equation}
At this point, we have changed the solution qualitatively. The introduction of eccentricity damping has added a complex component to the eigenfrequencies. Consequently, in the secular solution (6), real exponential decay factors appear in front of the oscillatory solution. The eigenvectors are now damped. Furthermore, the imaginary components of the eigenfrequencies need not be equal, and generally will not be, except for a narrow set of system parameters. This implies that the decay timescale of one of the modes,
\begin{equation}
\tau_{decay}^{(j)} = \left( \textrm{Im}\left[ g_j \right] \right)^{-1}
\end{equation}
can be considerably longer than all others, and the system will eventually evolve to a state that is characterized by a single eigenmode. Note that the eigenmode decay timescale can greatly exceed the tidal circularization timescale, prolonging the lifetime of the dissipated planets' eccentricities. Upon inspection of equation (6), it is clear that once the system is characterized by a single eigenmode, the rates of orbital precession are identical for all planets in the system. From Lagrange's planetary equations, this automatically implies that the apsidal angles between the orbits must be equal to $\Delta \varpi = 0$ or $\Delta \varpi = \pi$. In other words, all orbits are either aligned or anti-aligned, depending on which particular eigenmode has survived. Additionally, in this case, \textit{the ratios of the eccentricities are also well-defined by the eigenvector of the surviving mode}. When the system has reached a state where its dynamics are characterized by a single mode, it has reached a ``fixed point." Addition of higher-order terms to the disturbing function will modify the the eccentricity ratios implied by the fixed point, but will not cause $\Delta \varpi$ to be anything other than $0$ or $\pi$.

\subsection{Determination of $\sin(I)$}

We now have all the necessary ingredients to determine the system inclination. Usually, the dissipation time-scale greatly exceeds the secular time-scale ($\bf{C} \ll \bf{A},\bf{B}$), so  a system at a fixed point is characterized by a single eigenvector of the $\left[ \bf{A}+\bf{B} \right]$ matrix\footnote{The physical effect of including $\bf{C}$ in the solution is to offset the apses by a small factor proportional to $Q^{-1}$.}. From the definitions of the coefficients of $\bf{A}$ (equations 2 \& 3), it is clear that they are linearly proportional to $\sin(I)$. In fact, we can replace the true masses, $m$, by the minimum masses $\tilde{m}$ in equations (2 \& 3) and write $\bf{A} = \bf{\tilde{A}}/\sin(I) $. $\bf{B}$ is however independent of the system inclination, given that GR is the only contributing factor. Recall that this is the case for Neptune-sized and smaller planets, for which $\Lambda_{spin} \ll 1$ and $\Lambda_{tidal} \ll 1$. As a result, the eigenvectors of the $\left[\bf{\tilde{A}}/\sin(I)+\bf{B} \right]$ matrix, which physically correspond to the eccentricity ratios of the planets, depend explicitly on $\sin(I)$. Namely, every value of the system inclination corresponds to an eccentricity ratio of the planets. Consequently, \textit{a precise observational determination of the eccentricity ratios yields the true masses of the system}. Let us turn to an illustrative example below.

\subsection{Beyond Linear Order in $e$: \newline
 the Case of Well-Separated Orbits}

\begin{figure}[t]
\includegraphics[width=0.5\textwidth]{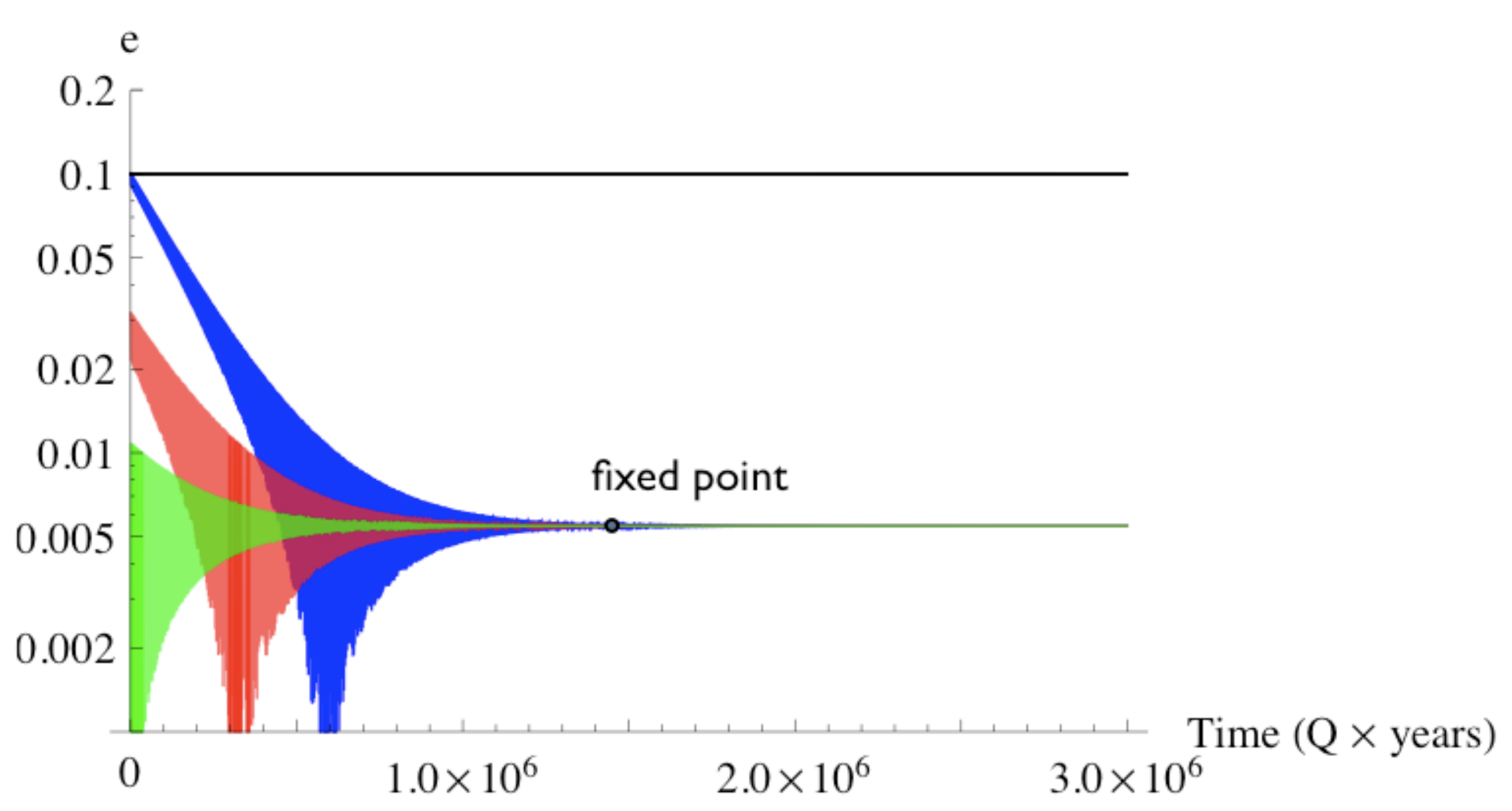}
\caption{A damped, modified Laplace-Lagrange secular solution of a 2-planet system with $m_1 = 10^{-5} M_{\odot}, m_2 = 10^{-2} M_{\odot}, a_1 = 0.03 \textrm{AU,} \ a_2 = 0.3 \textrm{AU} \ (\alpha = 0.1)$ and $e_2=0.1$. Three solutions are presented corresponding to the initial conditions $e_1=0.1$ (Blue), $e_1=0.03$ (Red), and $e_1 = 0$ (Green), with randomly chosen longitudes of perihelia. The black line shows the eccentricity of the outer planet. The apsidal angles initially circulate, but switch to libration at $t \approx Q \times 6 \times 10^5 $years. The system reaches a fixed point as the anti-aligned ($g_1$) mode decays away completely at $t \approx Q \times 1.3 \times 10^6 $years. Note that the system looses memory of its initial conditions as it approaches the fixed point.} 
\end{figure}

Consider the case of two well-separated ($\alpha \ll 1$) secondaries, where the inner planet is on a close-in orbit. In such a scenario, we only need to consider the additional apsidal precession of the inner planet. Since $\alpha \ll 1$, it is sensible to expand Laplace coefficients in equations (2-3) into hypergeometric series and retain only the first terms: $b_{3/2}^{(1)}(\alpha) \approx 3 \alpha$, $b_{3/2}^{(2)}(\alpha) \approx (15/4) \alpha^2$. With a little algebra, it is easy to show that to leading order in $\alpha$ and $\eta$, the eigenfrequencies take on a simple form:
\begin{equation}
g_{1}= \frac{3}{4}\frac{m_2 }{M_{\star} }n_{1} \alpha^3 (1+ \left[ \Gamma+ i\eta \right] ) 
\end{equation}
\begin{equation}
g_{2}= \frac{3}{4}\frac{m_1 }{M_{\star} }n_{2} \alpha^2 \left(1 + i \eta \left( \frac{5 \alpha }{4 (1+\Gamma )} \right)^2 \right)
\end{equation}
where $\Gamma \equiv B_{11}/A_{11}$ and $\eta \equiv C_{11}/A_{11}$. The two eigenfrequencies physically correspond to modes dominated by the inner ($g_1$) and outer ($g_2$) apsidal precessions. Note that the imaginary components of the modes have explicitly different dependences on $\alpha$. The multiplier in equation (16) is just ${A}_{11}$, expanded to first order in $\alpha$. So neglecting $\sin(I)$ for the moment, it is clear that $\textrm{Im}\left[ g_1 \right] = 1/ \tau_{c}$. This is consistent with the observation of Mardling (2007) that $\sim 3 \tau_c$ are needed for the system to attain a stationary state. The situation is wildly different however for the second mode, as $\textrm{Im}\left[ g_2 \right] = (25/16)(m_1/m_2)(\alpha^{5/2}/\tau_{c}) (1+\Gamma)^{-2}$. Consequently, equation (15) implies that $\tau_{decay}^{(2)} \gg \tau_{decay}^{(1)}$, unless $m_1 \gg m_2$ and the overall lifetime of the inner eccentricity is also greatly enhanced. 

\begin{figure}[t]
\includegraphics[width=0.45\textwidth]{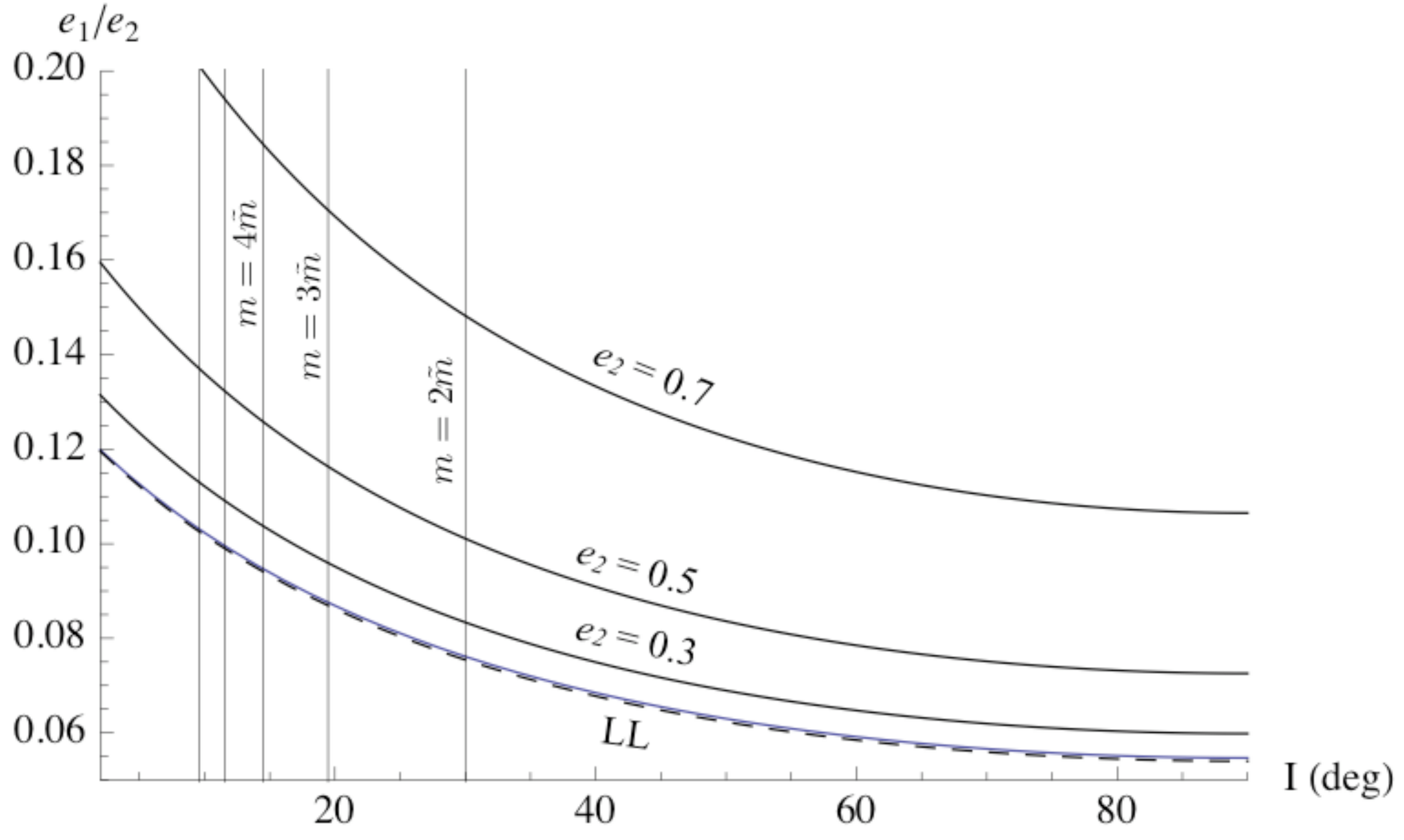}
\caption{Fixed point eccentricity ratio as a function of system inclination for a 2-planet system with $m_1 = 10^{-5} M_{\odot}, m_2 = 10^{-2} M_{\odot}, a_1 = 0.03 \textrm{AU and} \ a_2 = 0.3 \textrm{AU} \ (\alpha = 0.1)$ (see Fig. 1). The black curve, labeled LL was computed directly from the Laplace-Lagrange eigenvector solution. The blue dashed curve is the approximation to the LL solution, given by equation (19), corresponding to $e_2=0.1$. The curves with eccentricity labels demonstrate the dependence of the eccentricity ratios on the stationary eccentricity of the outer secondary, as dictated by the secular perturbation theory, developed by Mardling (2007). Recall that $\tilde{m}$ refers to the RV minimum mass.}
\end{figure}

The corresponding eigenvectors, also to leading order in $\alpha$, but neglecting the higher-order correction from $\eta$ read:
\begin{equation}
\left(\frac{\beta_{11}}{\beta_{12}}\right) = - \frac{4}{5 \alpha} \left(1- \frac{\tilde{m}_2}{\tilde{m}_1}\frac{1+\Gamma }{\sqrt{\alpha}}  \right) \gg 1
\end{equation}
\begin{equation}
\left(\frac{\beta_{21}}{\beta_{22}}\right) = \frac{5 \alpha}{4 (1+\Gamma )} \ll 1 
\end{equation}
Note that the eigenvector of the first mode is negative. By Euler's identity, the negative sign introduces an additional $i\pi$ in the exponent of the solution (6) for one of the planets. Physically, this corresponds to apsidal anti-alignment. Thus it is apparent from equations (18 \& 19) that the first and the second eigenmodes correspond to anti-aligned and aligned orbits respectively. 

As an illustration, consider a pair of planets with masses $m_1 = 10^{-5} M_{\odot}/\sin(I)$, $m_2 = 10^{-2} M_{\odot}/\sin(I)$, and semi-major axes $a_1=0.03 AU$, $a_2=0.3$ orbiting a $M_{\star}=1M_{\odot}$ star. The ($I=0$) damped, modified Laplace-Lagrange secular solution of this system is presented in figure (1), where the planets were started with $e_1 = e_2 = 0.1$ and randomly chosen longitudes of perihelia. The planetary Love number was chosen to be $k_{2p} =0.3$. Let us examine the evolution in some detail. After an initial transient period of $\sim 3 \tau_c$, the system reaches a fixed point. Thereafter, the free eccentricity decays on the timescale of  $\tau = \textrm{Im}\left[ g_2 \right] \approx 10^5 \tau_{c}$. As already stated, the addition of a perturbing planet has prolonged the lifetime of the dissipated planet's eccentricity immensely. As a result, it must be pointed out that the detection of an eccentric close-in planet alone does not imply that the planet itself is weakly dissipative. Rather, self-consistent calculations are required to place any constraints on $Q$.

The above analysis implies that planets on well-separated orbits in a tidally relaxed system will be apsidally aligned rather than anti-aligned, with the fixed-point eccentricity ratio, $e_1/e_2$, given by the corresponding eigenvector. Figure (2) shows the solution for the eccentricity ratio as a function of system inclination, $I$. The solid line, labeled LL, represents the directly calculated eigenvector and the dashed line represents the approximate solution, given by equation (19).

As can be inferred from figure (2), and equation (19), the fixed point eccentricity of the inner planet is much smaller than that of the outer planet. This is troublesome in the context of LL theory, where the outer eccentricity is already assumed to be small, because a precise observational determination of the eccentricity ratio becomes difficult. Consequently, we need to lift the constraint on the outer secondaries` eccentricity, so that the inner one at least becomes observably large. This can be accomplished by utilizing the secular perturbation theory, developed by Mardling (2007). The particular expansion of the disturbing function in terms of semi-major axes ratios places no restriction on the outer eccentricity in the equations of motion. Consequently, we can solve for the eccentricity ratio of the two planets by explicitly equating the precession rates of the two planets, given by  
\begin{eqnarray}
\frac{d \varpi_{1}}{dt} = \frac{3}{4} n_{1} \left( \frac{m_{2}}{M_{\star}} \right) \left( \frac{a_{1}}{a_{2}} \right)^{3}\frac{1}{(1- e_{2}^{2})^{3/2}} \times \nonumber \\ \left[ 1- \nu \frac{5}{4} \left( \frac{a_{1}}{a_{2}} \right) \left( \frac{e_{2}}{e_{1}} \right) \frac{1}{1- e_{2}^{2}} \right]+ B_{11}
\end{eqnarray}
\begin{eqnarray}
\frac{d \varpi_{2}}{dt}=\frac{3}{4} n_{2} \left( \frac{m_{1} }{M_{\star}} \right) \left( \frac{a_{1}}{a_{2}} \right)^{2} \frac{1}{(1- e_{2}^{2})^{2}} \times \nonumber \\ \left[ 1- \nu \frac{5}{4} \left( \frac{a_{1}}{a_{2}} \right)\left( \frac{e_{1}}{e_{2}} \right) \frac{1+4e_{2}^2}{1-e_{2}^2} \right]
\end{eqnarray}
where $\nu = \cos(\varpi_{1}-\varpi_{2})=\pm1$. These equations explicitly reveal that $(e_1/e_2)$ is not independent of $e_2$, as suggested by the eigenvector solutions. Note however, that the same expression for the eigenvectors (18)-(19) can be derived from these equations by expanding them to linear order in $e$ and solving for $(e_1/e_2)$. The solutions for $(e_1/e_2)$ as a function of $I$, obtained using equations (20) and (21) are also shown in figure (2) for various values of $e_2$. 

\begin{figure}[t]
\includegraphics[width=0.5\textwidth]{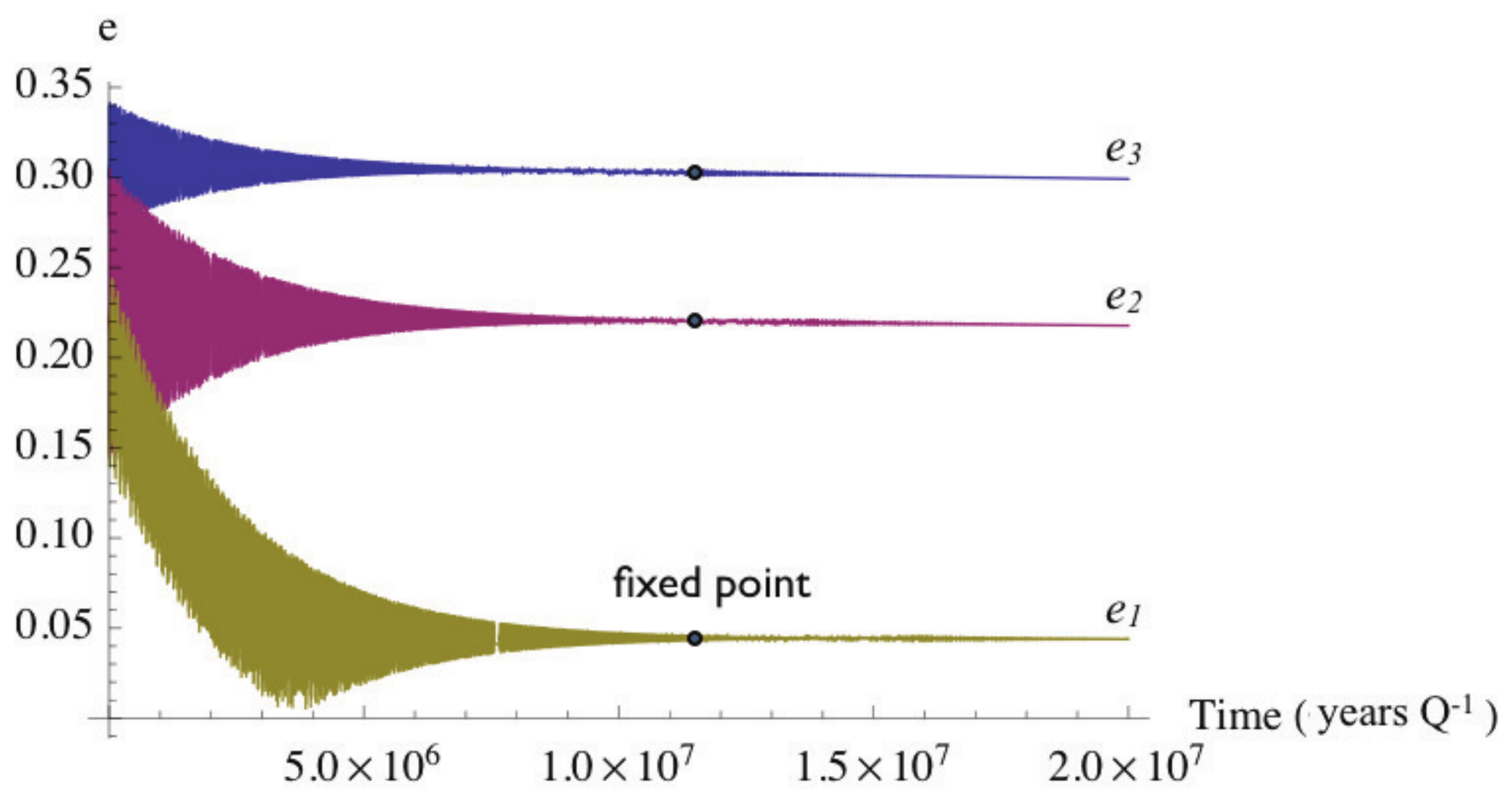}
\caption{A modified, dissipated Laplace-Lagrange secular solution of the 61 Vir system. The initial conditions were identical to those, listed in table (1).}
\end{figure}

\subsection{The General Case: Gauss's Averaging Method}
The above examples are illustrative in nature and are applicable when the appropriate assumptions are satisfied. It is also useful, however, to consider a general method that will be applicable in all cases, as long as the interactions among the planets are secular in nature. 

Rather than expanding the disturbing function in terms of a small parameter and applying Lagrange's planetary equations, consider N coplanar interacting elliptical wires of mass where the line density is inversely proportional to orbital speed and the integrated mass of the wire amounts to that of the planet (Gauss 1818). The magnitude of the force exerted on line element $r_j df_j$ by a line element $r_k df_k$ is simply
\begin{equation}
F_{jk}= G \frac{\rho_j \rho_k r_j r_k}{\Delta_{jk}^2}  df_j df_k  
\end{equation}
where $r$ is orbital radius, $\rho$ is density, $f$ is true anomaly, and $\Delta = | r_j - r_k |$ is the distance between the line elements. The radial and a tangential components of the force on line elements $j$ and $k$ are then
\begin{equation}
R_{jk}=F_{jk} \frac{r_k \cos(\phi)-r_j }{\Delta} \ \ \ \ \ T_{jk}=F\frac{r_k \sin(\phi) }{\Delta}
\end{equation}
\begin{equation}
R_{kj}=-F_{jk} \frac{r_k -r_j \cos(\phi)}{\Delta} \ \ \ \ \ T_{kj}=-F\frac{r_j \sin(\phi) }{\Delta}
\end{equation}
where $\phi = (f_k + \varpi_k - f_j - \varpi_j)$ is the angle between the line elements (Murray \& Dermott 1999). Recall that we are only interested in the situation where $\varpi_1 - \varpi_2 = (0,\pi)$. Following Burns (1976), the perturbation equation for the precession for longitude of perihelion reads
\begin{eqnarray}
\frac{d \varpi_j}{dt} &=&  \sqrt{\frac{a_j   (1-e_j^2) }{m^2_j  e^2_j G M_{\star} }} \oint \sum_{k=1, k \neq j}^N [ -\cos(f_j) \oint R_{jk} df_k \nonumber \\ &+& \frac{(2+e_j \cos{f_j}) \sin(f_j)  }{1+e_j \cos{f_j}} \oint T_{jk} df_k  ]  df_j + B_{jj}
\end{eqnarray}
 with an identical equation for $d\varpi_k/dt$. Note that in this formulation, as before, the secular term is linearly proportional to $\sin(I)$, unlike the GR correction. Thus, the system inclination can be solved for in the same way as above, but without constraints on eccentricity of semi-major axes.
 
\section{Application: 61VIR}
To date, the number of detected multi-planet systems that host small close-in planets remains limited to a handful of systems: HD 40307, 55 Cnc, 61 Vir, GJ 581 and GJ 876. Furthermore, the data for these systems are still comparatively sparse, so the error bars on the planet's eccentricities are rather large. These issues will surely get resolved with time, but at this point we can only give a rough assessment, and shall limit our analysis to a single case: 61 Vir.

The planetary system around the nearby sun-like star 61 Vir was discovered by Vogt et al (2010). The star hosts 3 planets, with orbital periods of roughly $4.2$d,  $38$d and $124$d (see Table 1 for an orbital fit). A simple evaluation of the system's dynamical stability yields no useful constraints on the inclination of the system. However, the minimum mass of the inner-most planet of $\tilde{m}=5.1 \pm 0.6 M_{\oplus}$ corresponds to that of a super-Earth, making it an ideal candidate for our method.
\\ 
\begin{figure}[t]
\includegraphics[width=0.5\textwidth]{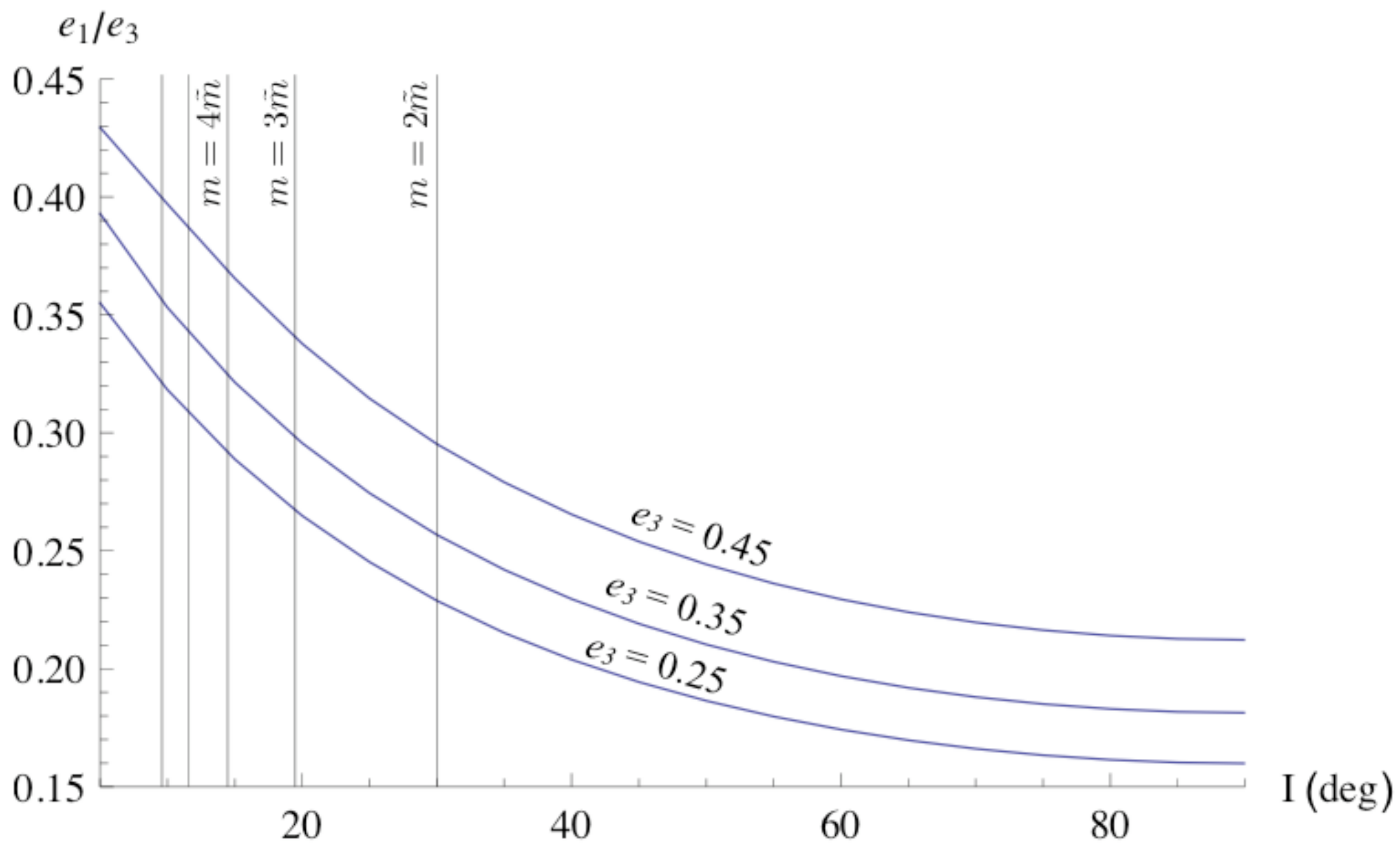}
\caption{Fixed point eccentricity ratio of planet b to planet d as a function of the 61 Vir system inclination. The curves were computed using Gaussian averaging method, with different stationary eccentricities of planet d, as labeled.}
\end{figure}
\\
The characteristic isolated circularization timescale of planet b is roughly $\tau_c \sim Q \times 10^6$ years. A damped, modified LL solution (shown in Figure 3) reveals that depending on starting conditions, up to 10 $\tau$ is required for the system to arrive to the fixed point. Thus, as already pointed out by Vogt et al (2010), given the star's multi-billion year age, we expect the system to be stationary if $Q_b \lesssim 10^3.$ For the illustrative purposes of this paper, we assume that planet b's tidal quality factor is similar to that of rocky bodies i.e. $Q_b = 100$.

Initially, we proceed as described in section 2.1 and compute the surviving LL eigenvector that physically corresponds to a state where all orbits are apsidally aligned. Given the moderate eccentricity ($e > 0.1$) of the outer two planets, however, the LL solution does not give a quantitatively acceptable answer. Consequently, we recompute the eccentricity ratios using the Gaussian averaging method, as described in section 2.4, utilizing the LL solution as an initial guess in the root-finding algorithm. The resulting curves are plotted in figures (4) and (5). It is noteworthy that although the Gaussian and LL solutions are qualitatively similar, higher-order secular terms clearly make a noticeable contribution to the fixed-point solution. 

Although the error bars on the orbital elements are still large, it is noteworthy that the observed system is consistent with a fixed point configuration. Thus, further observation of the system is warranted, given that if the system is found to be in a stationary state, it would yield not only the true masses, but also a constraint on the tidal quality factor of the inner-most planet. 
\\ 
\begin{figure}[t]
\includegraphics[width=0.5\textwidth]{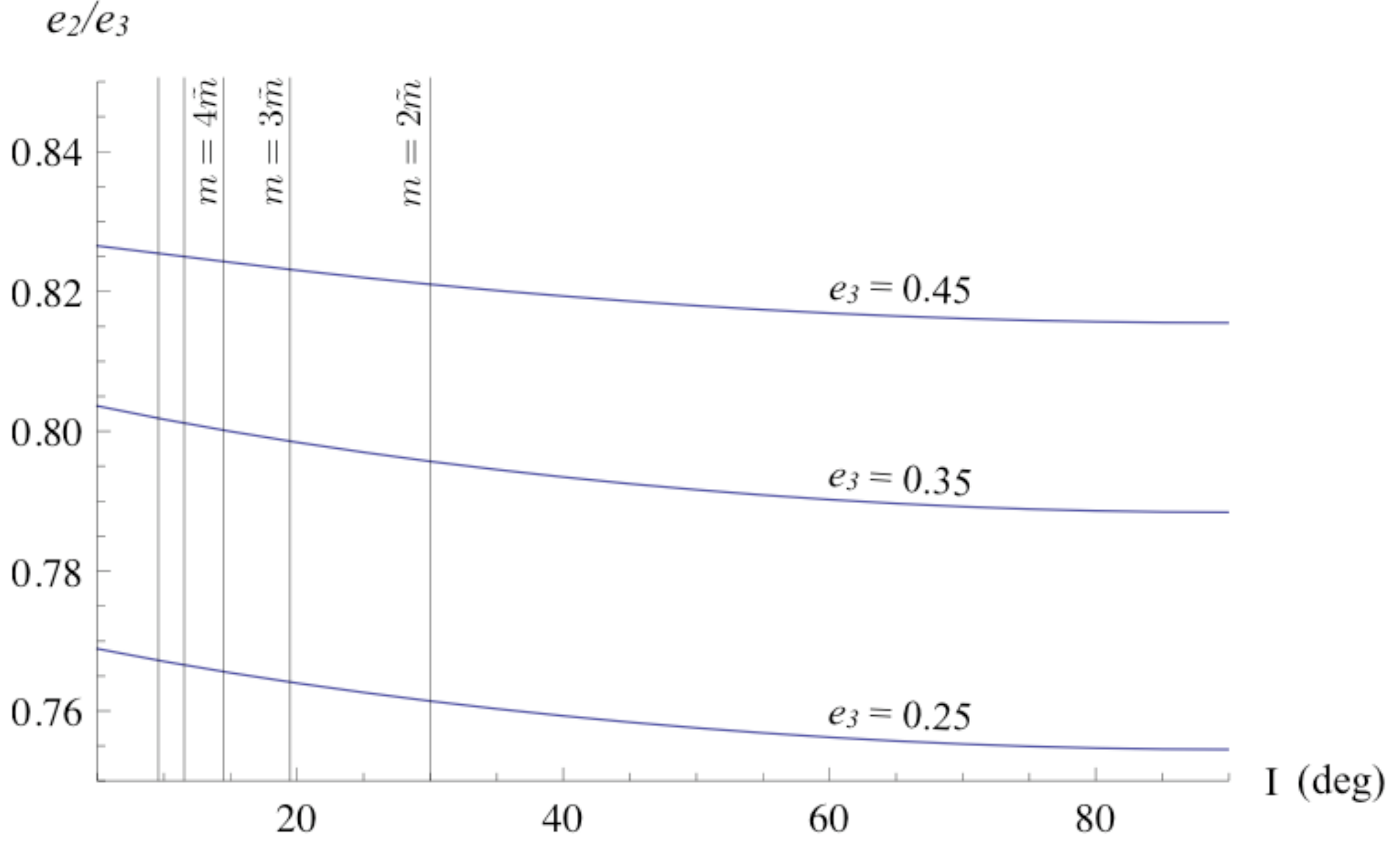}
\caption{Fixed point eccentricity ratio of planet c to planet d as a function of the 61 Vir system inclination. The curves were computed using Gaussian averaging method, with different stationary eccentricities of planet d, as labeled.}
\end{figure}
\\
\begin{table}
\begin{center}
\caption{Orbital Fit of the 61 Vir System}
\begin{tabular}{cccccccr}
\tableline \tableline
Planet & Mass ($m_{\oplus}$) & P (days) & e & $\varpi$ (deg) & \\
\tableline \\
b & 5.28 & 4.3 & 0.147 & 104 & \\
c & 19.1 & 38 & 0.155 & 331 & \\
d & 23.4 & 123 & 0.34 & 314 & 
\end{tabular}
\end{center}
\end{table}

\section{Comments on Massive Planets}

The domain of applicability of the method described in this paper does not extend to ``large" planets (recall that we require $\Lambda_{tidal}^{p} \ll 1$ in order to solve for $\sin(I)$). However, for massive, close-in planets, the $\sin(I)$ degeneracy can be resolved from spectral characterization of the host star alone (Snellen et al 2010). In such a case, the orbital precession rate yields information on the radius and the interior structure of the planet.

If only a single planet is present in the system, then the method described by Ragozzine \& Wolf (2009) can be employed. Namely, if the planet is sufficiently close to its host star, the orbital precession rate may be as high as a few degrees/year. In this case, direct observation of the orbital precession can be related to the sum of equations (7) -(9). As already discussed above, however, the first term in equation (9) dominates all other terms for large, massive planets. Consequently, $k_2(R)^5$ can be inferred. 

In order to accurately measure orbital precession, especially within the context of RV observations, significantly non-zero orbital eccentricity is needed. This poses a problem, since the eccentricities of single close-in planets are usually damped out on the timescale of $\sim 1$ Gyr. As a result, in practice, the method of Ragozzine \& Wolf (2009) is much better suited for transiting planets, where ultra-precise photometry, such as that characteristic of the $Kepler$ mission, can be used to pinpoint even a low ($\sim 10^{-3}$) eccentricity.

If there are two or more planets in the system, the situation is considerably more advantageous, since a finite eccentricity of the inner planet can be maintained over the age of the star by a perturbing companion. In this case, under the assumption of co-planar planets, the characterization of the fixed point through equations (20) and (21), where $B_{11}$ is dominated by the planetary tidal term, yields $k_2(R)^5$. In essence, the calculation is analogous to that of Batygin, Bodenheimer \& Laughlin (2009) for the Hat-P-13 system, with the exception that the radius is also unknown. Unfortunately, for a given mass, $k_2(R)^5$ is not a single-valued function of $R$ so the values of $k_2$ and $R$ cannot be disentangled by modeling of the planetary interior. 

Determination of $k_2 R^5$ of interest because the number of RV systems where the calculation is applicable is bound to greatly exceed the number of transiting systems for which $k_2$ can be measured directly, and a substantial distribution can be formed. The results of the \textit{Kepler} mission will provide a statistical distribution for planetary radii. However, because the majority of stars in the \textit{Kepler} field of view are faint, RV follow-up of most systems will be difficult. This poses a challenge for determination of $k_2$ by the method proposed by Batygin, Bodenheimer \& Laughlin (2009). Consequently, there is considerable value in deriving a statistical distribution for Love numbers from these observations.

\section{Conclusion}

\begin{figure}[t]
\includegraphics[width=0.5\textwidth]{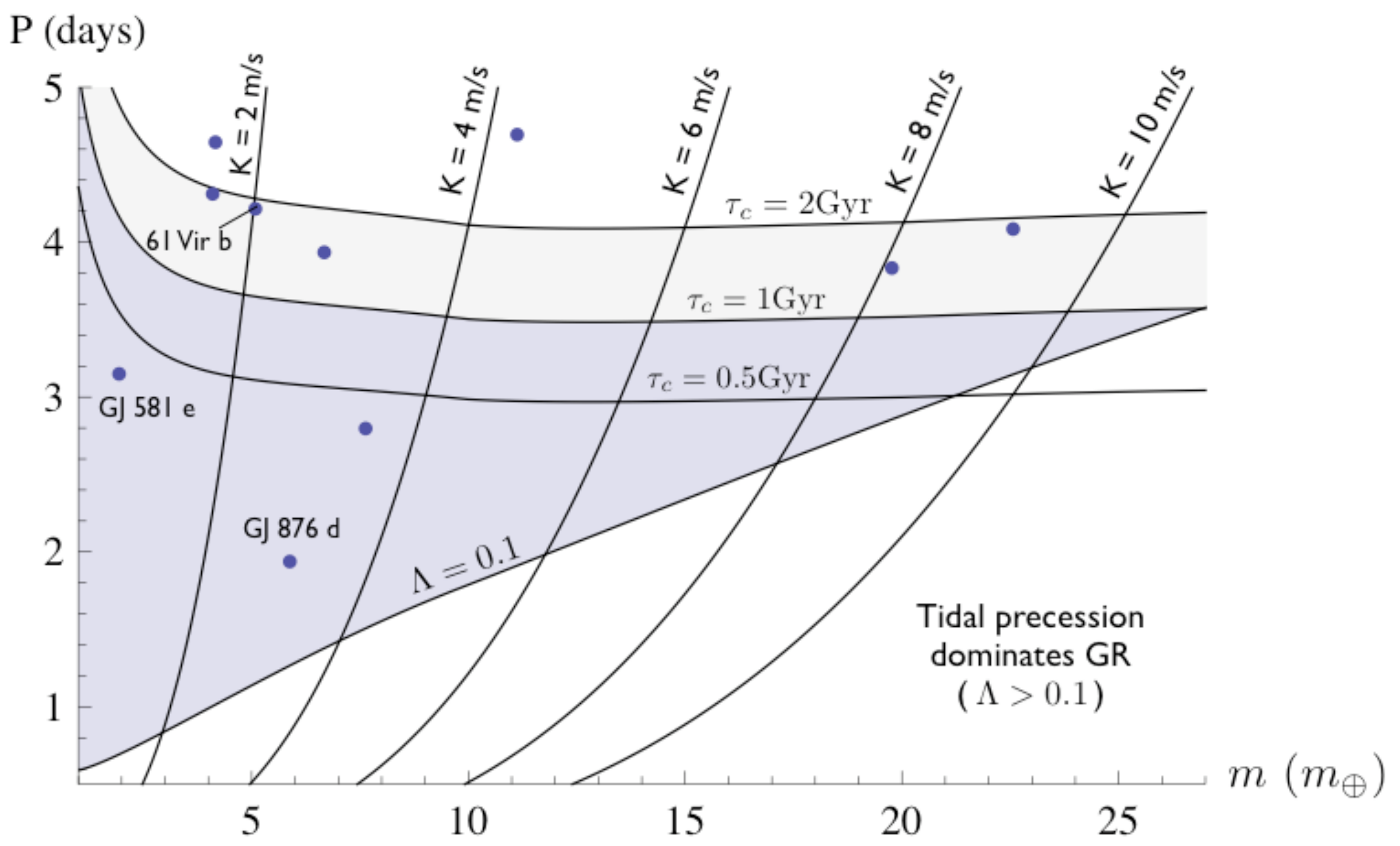}
\caption{Parameter space over which the method described here is applicable. The purple region is characterized by planets with circularization timescales less that 1 Gyr and non-relativistic effects contributing to less than 10\% of the non-secular precession. The blue dots correspond to currently known low-mass RV planets. Contours of RV signal semi-amplitudes are also shown. }
\end{figure}

In this paper, we present a method for determination of the true masses of RV planetary systems with a close-in planet. The analysis in question has important implications. First and foremost, it opens up a direct avenue towards an identification of the true lowest-mass exo-planets detected. This provides a direct constraint on the discussion of the habitability of RV planets. The second implication is more indirect. In a recent study, Ho \& Turner (2010) showed that there is significantly more uncertainty in $\sin(I)$ than previously assumed. In particular, the assumption that $\sin(I)$ has been drawn from a flat distribution is incorrect. Instead the distribution from which $\sin(I)$ is drawn is sensitively dependent on the true masses. Without additional information, it appears that there is significant adversity in estimating true masses of exoplanets from observations alone. As a result, resolution of $\sin(I)$ using an independent technique holds value not only in just yielding the true masses of a particular system, but also in implicitly constraining the relevant true-mass distribution from which $\sin(I)$ can then be drawn for the entire sample. 

It is certainly worthwhile to consider the observability of the systems to which our method is applicable. Recall that our method relies on three assumptions. First, tidal dissipation of orbital energy by the inner-most planet of a system must be efficient enough for the system to become tidally relaxed on a time-scale, less than a few Gyr i.e. the age of the star. Second, we require rough coplanarity\footnote{``Rough" coplanarity implies that the terms in the disturbing function that have the mutual inclination as a multiplier are small in comparison with terms of the same order that contain only the eccentricities.} of the system to ensure that fixed-point eccentricities are unaffected by the precession of the ascending node (Mardling 2010). Finally, to separate the dependence on $\sin(I)$ in the equations of motion, we require that the additional precession of the perihelion of the inner-most planet arises primarily from GR. Upon satisfaction of the above criteria, $\sin(I)$ can be solved for in an explicit, direct way. 

To demonstrate the extent of parameter space over which our method is applicable, we delineated the region where non-GR contributions account for less than 10\% of the additional precession of the inner-most planet, and circularization timescale is less than 1 Gyr.  Figure (6) shows this range, along with the current aggregate of low-mass RV planets. Given the uncertainty in tidal $Q$ as a function of planetary mass, $\tau_c = 2$ Gyr and $\tau_c = 0.5$ Gyr curves are also presented. Additionally, contours of corresponding semi-amplitudes of RV signal ($K$) are also displayed. Although the parameter space covered is considerable, it is clear that approximately 3-day period hot Neptunes make the best candidates for our method because of the optimum interplay between $K$ (making the planets most readily observable), and $\tau_c$. Finally, we discuss the possibility of obtaining information about the radius and interior structure of massive hot Jupiters in multiple systems, where the $\sin(I)$ degeneracy can be resolved with observations alone. Consequently, we encourage continued RV observation and more importantly, \textit{follow-up} of qualifying multi-planet systems, with the goal to pinpoint the orbital state to a high precision, thus deriving true masses and constraining the interior structure of low and high-mass RV exoplanets, respectively. 

We would like to conclude by presenting a list of possibilities for determination of physical properties of planets from observations of orbital parameters. The compiled flow-chart is presented as Figure 7. Let us summarize: if a newly discovered system harbors only a single tansiting hot Jupiter, the interior structure can be derived from monitoring of orbital precession (Ragozzine \& Wolf 2009). Alternatively, although observationally challenging, the rotational and tidal bulges can be deduced directly from the shape of the light-curve (Carter \& Winn 2010, Leconte et al 2011). If two planets are present and reside at a fixed point, the situation becomes more advantageous. If tidal precession plays an important role, and the inner planet transits, the Love number can be derived from a single snap-shot observation of the orbital state (Batygin Bodenheimer \& Laughlin 2009). If the inner planet does not transit, its exact mass can be derived spectroscopically (Snellen et al 2010) and $k_2 R^5$ can be computed. On the other hand, if GR overwhelms tidal precession, $\sin(I)$ degeneracy of the system can be resolved. If the system is tidally relaxed but is not co-planar, orbital evolution will follow a limit cycle rather than a fixed point (Mardling 2010).
\\
\\
{\footnotesize 
REFERENCES
\\ 
\\
Adams, F.~C., \& Laughlin, G.\ 2006, \apj, 649, 992 \\
Bakos, G.~{\'A}., et al.\ 2009, \apj, 707, 446 \\
Batygin, K., Bodenheimer, P., \& Laughlin, G.\ 2009, \apjl, 704,

L49 \\
Brouwer, D., \& van Woerkom, A.~J.~J.\ 1950, Astronomical 

papers prepared for the use of the American ephemeris and 

nautical almanac, v.13, pt.2, Washington : 

U.S.~Govt.~Print.~Off., 1950., p.~81-107 : 29 cm., 13, 81 \\
Burns, J.~A.\ 1976, American Journal of Physics, 44, 944 \\
Carter, J.~A., \& Winn, J.~N.\ 2010, \apj, 716, 850 \\
Eggleton, P.~P., \& Kiseleva-Eggleton, L.\ 2001, \apj, 562, 1012 \\
Gauss, C.~F.\ 1818, Gottingen : Heinrich Dieterich, 1818; 1 v.~; 

in 8.; DCCC.4.67 \\
Goldreich, P.\ 1963, \mnras, 126, 257 \\
Goldreich, P., \& Soter, S.\ 1966, Icarus, 5, 375 \\
Ho, S., \& Turner, E.~L.\ 2010, arXiv:1007.0245 \\
Laskar, J.\ 1986, \aap, 166, 349 \\
Laskar, J.\ 2008, Icarus, 196, 1 \\
Lagrange, J. L., \ 1776, Mem. Acad. Sci. Berlin, 199  \\
Laplace, P. S., \ 1772, M\'emoire sur les solutions particuli\`eres des 

\'equations diff\'erentielles et sur les in\'egalites s\'eculaies des plan\`ates 

Oeuvres completes, 9, 325 \\
Laughlin, G., Deming, D., Langton, J., Kasen, D., Vogt, S., Butler, P., Rivera, E., 
\& Meschiari, S.\ 2009, \nat, 457, 562\\
Leconte, J., Lai, D., \& Chabrier, G.\ 2011, arXiv:1101.2813
} \\
In this case, precise modeling can yield constraints on the mutual inclination between planets. If three or more planets are present in the system, the solution simplifies to one that is similar to the two-planet case if the system is at a fixed point. Otherwise, the situation is considerably more complex and should be treated on a case-by-case basis.  Finally, it is always important to keep in mind that measurement of flux-excess during secondary eclipse can yield the tidal luminosity of a planet (e.g. Laughlin et al 2009), and thus its tidal Q.  

\begin{figure}[t]
\includegraphics[width=0.8\textwidth]{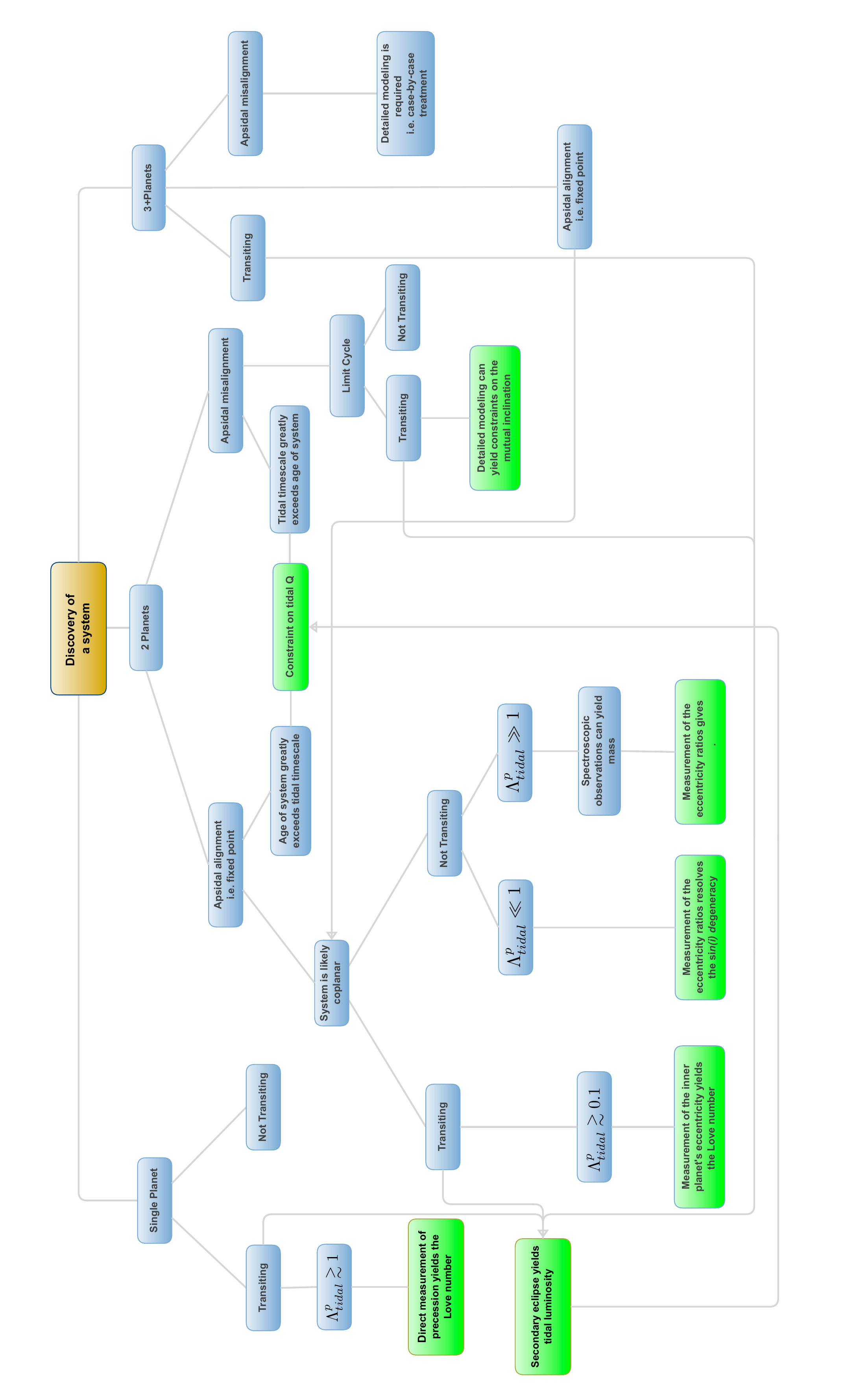}
\caption{A flow-chart that depicts various possibilities for determination of physical properties of planets from observations of their orbital configurations.}
\end{figure}

\acknowledgments

We are grateful to Y. Wu, D. J. Stevenson, J. A. Johnson and M. E. Brown for carefully reviewing the manuscript and for useful discussions. We thank the anonymous referee for useful suggestions. This work was funded by NASA grant NNX08AY38A and NASA/Spitzer/JPL grant 1368434 to GL.
\\
\\
\\
{\footnotesize
Le verrier, U. J. J., \ 1856, Ann. Obs. Paris. II \\
Lee, M.~H., \& Peale, S.~J.\ 2003, \apj, 597, 644 \\
Lo Curto, G., et al.\ 2010, \aap, 512, A48 \\
Lovis, C., et al.\ 2010, arXiv:1011.4994 \\
Mardling, R.~A., \& Lin, D.~N.~C.\ 2002, \apj, 573, 829 \\
Mardling, R.~A.\ 2007, \mnras, 382, 1768 \\
Mardling, R.~A.\ 2010, \mnras, 407, 1048 \\
Mayor, M., \& Queloz, D.\ 1995, \nat, 378, 355 \\
Morbidelli, A.\ 2002, Modern celestial mechanics : aspects of 

solar system dynamics, by Alessandro Morbidelli.~London: Taylor Francis, 2002 \\
Michtchenko, T.~A., \& Malhotra, R.\ 2004, Icarus, 168, 237 \\
Migaszewski, C., \& Go{\'z}dziewski, K.\ 2009, \mnras, 392, 2  \\
Murray, C.~D., \& Dermott, S.~F.\ 1999, Solar system dynamics by Murray, C.~D., 1999 \\
Ragozzine, D., \& Holman, M.~J.\ 2010, arXiv:1006.3727 \\
Ragozzine, D., \& Wolf, A.~S.\ 2009, \apj, 698, 1778 \\
Rivera, E.~J., et al.\ 2005, \apj, 634, 625  \\
Snellen, I.~A.~G., de Kok, R.~J., de Mooij, E.~J.~W., \& Albrecht, 

S.\ 2010, \nat, 465, 1049 \\
Sterne, T.~E.\ 1939, \mnras, 99, 451  \\
Terquem, C., \& Papaloizou, J.~C.~B.\ 2007, \apj, 654, 1110 \\
Vogt, S.~S., et al.\ 2010, \apj, 708, 1366 \\
Veras, D., \& Armitage, P.~J.\ 2007, \apj, 661, 1311 \\
 Wu, Y., \& Goldreich, P.\ 2002, \apj, 564, 1024 }

\end{document}